\documentclass{article}
\usepackage[centertags]{amsmath}
\usepackage{amssymb}
\usepackage{texdraw}
\usepackage{a4wide}
\usepackage[dvips]{epsfig}
\usepackage{pstricks}
\usepackage[dvips]{graphicx}
\newcommand{\dsp}{\displaystyle}
\newcommand{\be}{\begin{equation}}
\newcommand{\ee}{\end{equation}}
\newcommand{\bea}{\begin{eqnarray}}
\newcommand{\eea}{\end{eqnarray}}
\newcommand{\bi}{\begin{itemize}}
\newcommand{\ei}{\end{itemize}}

\input texdraw
\def\today{\number\day\ \ifcase\month\or January\or February\or
March\or April\or May\or June\or July\or August\or September\or
October\or November\or December\fi\ \number\year}
\numberwithin{equation}{section}
\font\bbigbf=cmbx12 scaled \magstep 1 
\font\bigsc=cmcsc10 scaled \magstep 1 \font\bigrm=cmr12
\font\filt=msbm10 \font\filtsv=msbm10 scaled 700

\def\reals{\hbox{\filt R}}
\def\svreals{\hbox{\filtsv R}}

\def\re{\mathop{\rm Re}\nolimits}
\def\im{\mathop{\rm Im}\nolimits}
\def\arcsinh{\mathop{\rm arcsinh}\nolimits}
\def\bv#1{\hbox{$\mathbf{#1}$}}
\def\rs{r_\circ}

\def\pv{\;\mathop{\rm PV}\nolimits\!\!}
\def\Bs{{\mathcal{B}_\circ}}
\newcommand{\setlinespacing}[1]%
           {\setlength{\baselineskip}{#1 \defbaselineskip}}

\begin{document}
\begin{titlepage}
\null\bigskip \centerline{\bbigbf Field-induced motion of nematic
disclinations}
\bigskip\bigskip\bigskip\bigskip
\centerline{\bigrm {\bigsc Paolo Biscari}$^1$ and {\bigsc Timothy
J.\ Sluckin}$^2$}
\bigskip\bigskip\noindent
\small $^1$ Dipartimento di Matematica, Politecnico, Piazza
Leonardo da Vinci 32, 20133 Milano (Italy) and \\
Istituto Nazionale di Fisica della Materia. Via Ferrata 1 -- 27100
Pavia (Italy). \\
$^2$ Faculty of Mathematical Studies, University,
Southampton SO17 1BJ (United Kingdom). \normalsize

\vskip 1cm

\noindent Date: \today

\vskip 1 cm

\noindent 2000 MSC: 76A15 -- Liquid crystals

\vskip 1cm

\begin{abstract}
An individual defect in a nematic liquid crystal moves not only in
response to  its interaction with other defects but also in
response to an external  field.  We analyze the motion of a wedge
disclination in the presence of an applied field of strength $H$.
We neglect backflow and seek steadily travelling patterns. The
stationary picture yields a semi-infinite wall of strength $\pi$,
bounded by the defect line. We find that the disclination advances
into the region containing the wall at velocity $v(H)$, where $v$
scales as $H/|\log H|$ as long as the magnetic coherence length is
greater than the core radius. When the external field is applied
in the presence of a pair of disclinations, their dynamics is
strongly influenced. We compute the expected relative velocity of
the disclinations as a function of distance and field. The natural
tendency for the disclinations to annihilate each other can be
overcome by a sufficiently strong field suitably directed.
\end{abstract}
\end{titlepage}

\section{Introduction}

Singularities in  liquid crystals, or \emph{defects\/}, have
played a fundamental role  not only in  the development of the
understanding of the physics of liquid crystals but also in the
later development of the topological theory of defects in
condensed matter. In nematic liquid crystals, point, line and wall
defects can be found.  Line defects were first classified by F.C.\
Frank, who noted that line defects came in classes  with an
integer or half-integer  charge \cite{58fr}. Later workers, using
the topological theory of defects, realised that nematic liquid
crystals sustain a topologically unique line defect
\cite{79merm,89klem,97bigu}. However, Frank's na\"{\i}ve
classification, which effectively supposes that the nematic order
parameter is restricted to  a plane, remains important in
providing guidance and intuition for the physics of defects in
nematic liquid crystals.

The topological total charge of a system is conserved during its
evolution. Under many circumstances this remains true for Frank's
definition of charge; this is a stronger condition.  For instance,
defects of opposite charges may annihilate and defects of higher
topological charges may decay to a bunch of defects of smaller
topological charges. Topological dipoles may even be nucleated
from a smooth field \cite{99guvi}.

In this paper we focus on  nematic disclinations, \emph{i.e.\/}
line defects. According to Frank's definition, the topological
charge of a line defect is defined as the number of turns the
director performs  along a closed path  surrounding the defect.
This number may be half-integer because of the head-and-tail
symmetry of nematic liquid crystal molecules. When the final
director is rotated by an angle $\pi$ with respect to the initial
director, the physical state they describe does not exhibit any
discontinuity.

The physical manifestation of the topological theory of defects in
nematic liquid crystals concerns \emph{escape into the third
dimension\/}. Nematic disclinations of integer charge are not
topologically stable. A suitable continuous transformation,
involving the third previously neglected  dimension, dissolves the
singularity and leaves behind a regular field \cite{72clkl}. In a
similar manner, all half-integral disclinations can be distorted
into each other. But often free energy criteria impede this
process, and the Frank classification remains useful. In
particular, annihilation of opposite charge half-integral defects
is favored, but of same-charge defects is hindered.

Defect dynamics has been widely studied either in the classical
Oseen-Zocher-Frank (OZF) theory \cite{33osee,33zoch,58fr}, or
within the extended de Gennes-Ericksen theory
\cite{93dgpr,89eric}. However, there are fundamental dilemmas in
treating defect motion. The  OZF theory is not a dynamical theory,
and defect cores are regions in which, strictly speaking, the OZF
theory does not apply. The theory can be extended to deal with
moving defects, but only by introducing a phenomenological
dissipation function using director rate of change. This minimal
extension is incorrect; the full extension involves hydrodynamic
terms which depend on the local director \cite{92lesl}. But the
local director is not defined in the defect region, and so this
extension is also inadequate to describe defect motion. The
alternative de Gennes-Ericksen approach (now usually called the
\bv{Q}-tensor theory) is in principle up to the task, but now the
defect regions are no longer special; there is a danger of
confusing calculation with computation, leading to answers of
doubtful physical significance.

Despite these problems, some theoretical progress has been made,
in part because some  authors have detected analogous topological
structures in liquid crystals and cosmological models. A single
disclination may move through an otherwise smooth field
\cite{91rykr}, but then the problem is why the defect is moving in
the first place. An implicit  response to this question is to give
more attention to the interaction between two or more defects. Now
the defects move, slowly, to reduce their elastic energy. For
example,  attraction between two point defects of opposite charges
has been studied both in planar \cite{91patu,92piru,96denn} and
cylindrical \cite{96guvi,98guvi,99biguvi} geometries.   The
attraction between a dipole of smectic disclinations has been
studied in \cite{92pafi}.

More precise quantitative descriptions of the defect evolution
must necessarily take into account backflow effects
\cite{78clle,90olgo,92olgo}, \emph{i.e.} the interactions between
director rotation and macroscopic molecular motion. The first
analytical attempt to introduce backflow effects was performed in
\cite{00rich}, where the macroscopic velocity field was coupled to
the degree of orientation, though not to the director field. A
series of recent numerical simulations
\cite{00deor,01deor,02svzu,02tode} have determined the main
effects of backflow coupling. The dynamical director patterns turn
out to be strongly influenced in the final part of the
annihilation process, but some effects may be noted even during
all the defect evolution. In particular, the positive-charge
disclination moves almost twice as faster than the negative-charge
one \cite{02svzu}.

In this paper we investigate the effect of applied external field
on the defect dynamics. We consider a simple geometry: a single
$+\frac{1}{2}$, or a dipole of $\pm\frac{1}{2}$ disclinations, in
planar symmetry. We compute the defect speed with the aid of the
Leslie's dissipation balance \cite{92lesl}. It turns out that the
external field exerts a profound effect on the defect dynamics and
the defect interaction. By suitably adjusting the external field
strength and direction it is possible to drive a single
disclination through the sample, almost at will. More
interestingly, the external field may not only accelerate the
annihilation process. It can also stop the defects, or even
reverse their  velocities, so transforming an attractive into a
repulsive interaction. Depending on the external field direction,
a critical defect distance may arise. This characterizes the
defect interaction. They annihilate each other if they come closer
than that distance; otherwise, they repel.

Throughout our presentation we will introduce and discuss some
assumptions that simplify our analytical calculations. The
1-constant approximation in the elastic free energy, and a
parabolic approximation in the magnetic energy linearize the free
energy derivatives. We remark that the parabolic approximation
should be abandoned if we were to generalize the present study to
highly charged defects. We also neglect backflow. This
approximation allows us to derive an analytical condition which
determines the defect velocity, and in particular the critical
distance that reverses the defect interaction. A numerical
analysis would correct these computed values, even if the
described phenomena will certainly remain.

The plan of the paper is as follows. In Section 2 we analyse the
motion of a single disclination in an external field. Section 3
describes the defect interaction, and how it can be reversed with
the aid of the applied field. Section 4 briefly summarizes and
discusses our results.

\section{Single defect motion}

We consider a $+\frac{1}{2}$ disclination embedded in an external
magnetic field. This field will favor the director orientation of
one side of the defect with respect to the other side. This
asymmetry  is sufficient, as we shall see, for energy
considerations alone to  determine that the defect will move, and
to determine the direction of its motion. Our task is to determine
the magnitude of the velocity as a function of the field
intensity. We adopt the 1-constant approximation and neglect
backflow. This latter approximation implies that our estimated
velocity will be certainly smaller than the actual velocity. In
fact, backflow effects reduce the total dissipation, so allowing
faster director dynamics.

\begin{figure}[htb]
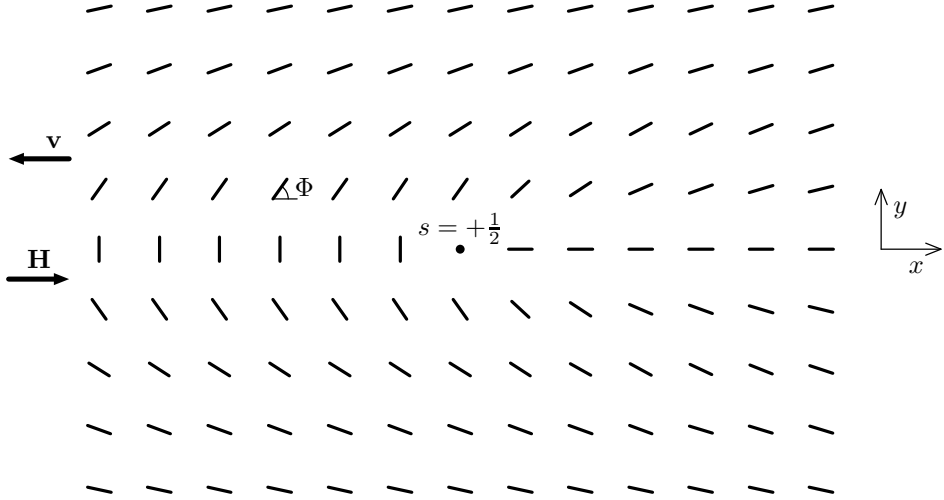

\null\medskip
\input fig1.tex
\caption{Geometric setting for the analysis of the single defect
motion. The defect occupies the origin of the co-moving reference
frame, and its moves towards the $\pi$-wall. The angle $\Phi$ is
determined by the director and the motion direction (orthogonal to
the external field).} \label{singdef}
\end{figure}

Let us consider the reference frame illustrated in Figure
\ref{singdef}. It moves with the disclination, which sits at the
origin $O$. The $x$-axis is parallel to the external field. Given
any point $P$ in the plane, let $\vartheta$ be the polar angle
between the radial direction $(P-O)$ and the $x$-axis.
Disclinations of half-integer charge do not benefit from escaping
into the third dimension. Thus, we restrict our interest to
situation in which the director \bv{n} is confined to the plane,
and let $\Phi(P)$ be the angle the nematic director at $P$
determines with the $x$-axis.

In the absence of a field, the 1-constant approximation implies
that the director angle varies linearly with the polar angle:
$\Phi=\frac{1}{2}\,\vartheta$. In the presence of the field the
same linear dependence holds, roughly speaking, on length scales
smaller than the magnetic coherence length $\xi$. So long as we
confine ourselves to these length scales, the elastic energy
overwhelms the external field. However, far from the defect, the
magnetic energy forces the system toward its preferred value
$\Phi=0$.  On the other hand,  over a closed path around the
defect, no matter how far that path may be from the defect,
topology \emph{forces} the angle $\Phi$ to rotate through $\pi$.
This creates a dilemma, for the system must rotate through an
angle $\pi$, and yet remain at $\Phi=0$.

These two constraints are reconciled by restricting the region
over which $\Phi$ rotates. This region is a topologically
irreducible \emph{wall} between regions of space whose director is
oppositely oriented. Associated with the wall is a well-defined
surface free energy analogous to the surface tensions of fluid
mechanics.  This is a $\pi$-wall, since the director concentrates
its $\pi$-rotation in it. In the presence of the field, the
previously isolated defect line has been transformed into the
trailing edge of a wall defect.  The disclination then moves into
the wall in order to reduce the wall area and consequently the
free energy of the system.

\subsection{Dissipation principle}

We determine both the director field and the value of the
disclination velocity by imposing the energy balance between the
free-energy loss-rate and the dissipation \cite{92lesl}:
\begin{equation}
\mathcal{\dot F}+\mathcal{D}=0\;. \label{dispr}
\end{equation}
In the 1-constant approximation, the free-energy functional is
given by
\begin{equation}
\mathcal{F}[\Phi] = \int_{\svreals^2} \left(K\big|\nabla
\Phi\big|^2+ \chi_{\rm a}H^2\sin^2\Phi\right)da\;, \label{inif}
\end{equation}
where $K$ is an average elastic constant, $\chi_{\rm a}$ is the
magnetic anisotropy, and $H$ is the strength of an external
magnetic field.

It is well-known that a $+\frac{1}{2}$ disclination possesses an
infinite core energy. The elastic free-energy density diverges at
the defect, and the singularity is not integrable. There are two
ways to avoid this divergence. The first consists in excluding
from our integration region a small disk centered in the defect,
the \emph{core region\/} $\Bs$. The radius of the excised disk,
the \emph{core radius\/} $\rs$, is usually much smaller than all
other characteristic lengths entering the problem, so that many
studies have been performed in the limit of vanishing $\rs$
\cite{86brcoli}. A more precise physical description of the defect
requires an extension of the classical Oseen-Zocher-Frank theory,
and the replacement of the nematic director with the nematic order
tensor \cite{87shsl}. We will choose the first option, and perform
all the integrations in the pierced domain, which excludes $\Bs$.
We further assume that the core radius is fixed. The basic physics
of the phenomenon is well-described using these approximations.
However, in order to deal with external fields of any intensity,
it would be interesting to apply the techniques of \cite{03bisl}
to determine how and when the magnetic coherent length influences
the core radius.

A second, though less worrying, free-energy divergence comes from
the supposedly infinite size of the domain. The domain may extend
indefinitely in the $y$-direction without inducing any
singularity, since both the elastic and the magnetic energy
densities vanish away from the $x$-axis. On the contrary, there
will be few cases in which a large-$x$ cut will be needed to keep
energies finite. In those cases we will assume that our domain is
bounded by $|x| \leq L$. Whenever possible, we will perform the
$L\to+\infty$ limit, and we will notice that the large scale
length $L$ will not finally enter in the defect velocity.

Our final approximation concerns the magnetic free-energy. We will
perform the \emph{parabolic approximation\/} $\sin^2\Phi\simeq
\Phi^2$. This approximation allows us to obtain linear field
equations in $\Phi$. It can be used in the whole domain provided
we define $\Phi\in[-\frac{\pi}{2},\frac{\pi}{2})$, since it that
case all the values attained by the director angle belong to the
potential well of the equilibrium configuration $\Phi=0$. We
remark that this approximation would not be valid if we were
interested in analysing the equilibrium configurations of more
complex defects.

When we neglect backflow, the dissipation assumes the simple
expression
\begin{equation}
\mathcal{D} = \gamma_1 \int_\Bs \dot\Phi^2\,dx\,dy\;,
\label{dissdef}
\end{equation}
where the pierced integration domain $\Bs$ comes into play since
also the dissipation density diverges in the core region.

Let us perform an infinitesimal displacement $\delta\Phi$ of the
director field. If we make use of the divergence theorem, the
dissipation principle (\ref{dispr}) requires
\begin{equation}
\int_{\partial \Bs} \delta\Phi\; \big(2K\,\nabla\Phi\big)\cdot
\mathbf{\nu}\;d\ell+\int_\Bs\delta\Phi\,\left[\gamma_1\,\dot\Phi
-2K\Delta\Phi+ 2\chi_{\rm a}H^2\Phi\right] da=0\qquad
\forall\:\delta\Phi\;,
\label{local0}
\end{equation}
where $\nu$ is the outer normal along $\partial\Bs$. The
arbitrariness of $\delta\Phi$ requires that the quantity in square
brackets in the second integral must vanish identically:
\begin{equation}
\gamma_1\,\dot\Phi=2K\Delta\Phi - 2\chi_{\rm a}H^2\,\Phi\;.
\label{eq0}
\end{equation}
Equation (\ref{eq0}) is the well-known time-dependent
Ginzburg-Landau evolution equation of the system. It will supply
us the director field. However, the Ginzburg-Landau equation alone
is not able to guarantee the dissipation principle. We must check
that also the first integral in (\ref{local0}) vanishes. The
boundary $\partial\Bs$ is made of two parts: a small circle around
the defect and a large boundary at infinity. The integral around
the latter vanishes since at infinity $\nabla\Phi$ fades
everywhere, except along the $\pi$-wall, where it is orthogonal to
the outward normal. We are then left with the first integral in
(\ref{local0}), performed along the boundary of the core region.
This quantity has a simple physical meaning \cite{02cefr}: it is
the power supplied by the core region to the rest of the domain.
Thus, to require that all the free-energy loss is dissipated
within the system is equivalent to require that no energy is
supplied to it, neither from the outside nor from the core region.

\subsection{Steadily moving defects}

We will look for stationary solutions of equation (\ref{eq0}).
They aim at representing a defect moving at a constant speed $v$
towards the $\pi$-wall. We will find that, for any positive value
of $v$, it is possible to find a solution of (\ref{eq0}) which
satisfies the boundary conditions. However, there is just one
value of the velocity which cancels also the first integral in
(\ref{local0}) (or, equivalently, that satisfies the global
dissipation condition). It is the only velocity at which the
defect is able to move without any external boost.

In a steadily moving reference frame, traveling with the defect
itself, the relative stationary differential equation follows by
replacing in (\ref{eq0}) the time derivative with $v\,\partial_x$.
We thus obtain:
\begin{equation}
\Delta\Phi -\frac{\lambda}{\xi}\;\frac{\partial\Phi}{\partial x} -
\frac{\Phi}{\xi^2} =0\;,
\label{difeq}
\end{equation}
where $\xi:=\dsp\sqrt\frac{K}{\chi_{\rm a} H^2}\,$ is the magnetic
coherence length, and
$\lambda:=\dsp\frac{\gamma_1}{2\sqrt{K\chi_{\rm
a}}}\,\frac{v}{|H|}$ is a dimensionless quantity. We look for
solutions of the eigenvalue problem (\ref{difeq}) that satisfy the
symmetry requirement $\Phi(x,-y)=-\Phi(x,y)$ for all $y\neq 0$,
and the boundary conditions
\begin{equation}
\Phi(x,0^+)=
\begin{cases}
0 & {\rm if}\ x>0 \\ \noalign{\smallskip} \dsp\frac{\pi}{2} & {\rm
if}\ x<0\,,
\end{cases}
\qquad{\rm and}\qquad \lim_{y\to\infty}\Phi(x,y)=0\quad {\rm for\
all}\  x\in\reals\;,
\label{bcond1}
\end{equation}
where the latter condition is determined by the presence of the
magnetic field. The boundary conditions (\ref{bcond1}) are
singular only at the origin, where a disclination of topological
charge $+\frac{1}{2}$ stands. Indeed, the discontinuity the angle
$\Phi$ suffers along the negative $x$-axis does not induce any
physical singularity, since $\Phi=\pi/2$ and $\Phi=-\pi/2$
describe the same director orientation.

Among the solutions of the eigenvalue problem
(\ref{difeq})-(\ref{bcond1}), the dissipation principle
(\ref{dispr}) will single out the only physical one.

\subsection{Director field}

\begin{figure}[htb]
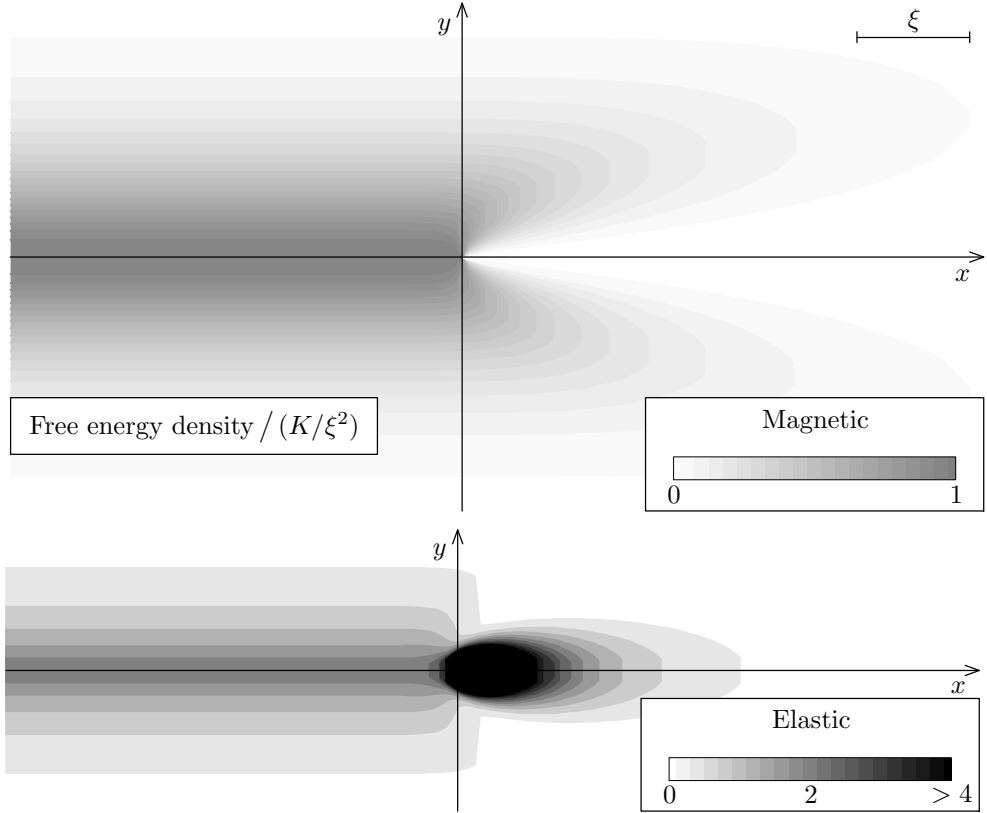

\begin{center}
\medskip
\input fig2a.tex

\medskip

\input fig2b.tex
\medskip
\end{center}
\medskip
\caption{Density plots of the magnetic (upper graph) and elastic
(lower) parts of the free-energy density, computed from the
analytical solution (\ref{ansol}). The elastic energy is localized
close to the defect. On the contrary, the magnetic energy is
mostly packed around the $\pi$-wall. The defect core is not
symmetric with respect to the defect position. It is slightly
shifted behind it, with respect to the direction of motion.}
\label{plotfield}
\end{figure}

We solve the linear partial differential equation (\ref{difeq})
with a Fourier transform. To this aim, it is useful to write the
first of the boundary conditions (\ref{bcond1}) as
\begin{equation}
\Phi(x,0^+)=\frac{\pi}{4}-\frac{1}{4{\rm i}} \pv\int_{\svreals}
\frac{{\rm e}^{{\rm i}qx}}{q}\,dq\qquad{\rm for \ all}\quad x\neq
0\;,
\label{bcond}
\end{equation}
where $\pv\;$ denotes the Cauchy Principal Value of an integral.
We then look for solutions of equation (\ref{difeq}) of the form:
\begin{equation}
\Phi(x,y)=\frac{\pi}{4}\,g_1(y)-\frac{1}{4{\rm i}}
\pv\int_{\svreals} \frac{{\rm e}^{{\rm i}qx}}{q}\,g_2(q,y)\,dq\;,
\label{ans}
\end{equation}
with
\begin{equation}
g_1(0)=g_2(q,0)=1\quad {\rm and}\quad
\lim_{y\to\infty}g_1(y)=\lim_{y\to\infty}g_2(q,y)=0
\quad{\rm for\ all}\ q\in\reals\,.
\label{bcondg}
\end{equation}
If we insert (\ref{ans}) in (\ref{difeq}) we obtain:
\begin{equation}
\frac{\pi}{4}\left(g''_1-\frac{1}{\xi^2}\,g_1\right)
-\frac{1}{4{\rm i}}\pv \int_{\svreals} \frac{{\rm e}^{{\rm
i}qx}}{q}\left[\frac{\partial^2 g_2}{\partial y^2}-
k^2(q)\,g_2\right] \,dq=0\,,
\label{anseq}
\end{equation}
where $k(q)$ will henceforth denote the positive-real-part
square-root of $k^2(q)=q^2 +\dsp\frac{{\rm
i}q\lambda}{\xi}+\frac{1}{\xi^2}\;$. The solution of (\ref{anseq})
and (\ref{bcondg}) in the upper half-plane $\{y\geq0\}$ is
\begin{equation}
\Phi(x,y)=\frac{\pi}{4}\,{\rm e}^{-\frac{y}{\xi}}-\frac{1}{4{\rm
i}} \pv\int_{\svreals} \frac{{\rm e}^{{\rm i}qx-k(q)y}}{q}\,dq\;,
\end{equation}
whereas, by symmetry, the solution in the whole plane is given by:
\begin{equation}
\Phi(x,y)=\varepsilon(y)\left[\frac{\pi}{4}\,{\rm
e}^{-\frac{|y|}{\xi}}-\frac{1}{4{\rm i}} \pv\int_{\svreals}
\frac{{\rm e}^{{\rm i}qx-k(q)|y|}}{q}\,dq\right]\;,
\label{ansol}
\end{equation}
with
\begin{equation}
\varepsilon(y):=
\begin{cases}
\phantom{-}1 & {\rm if}\ y\geq 0\;, \\
-1 & {\rm if}\ y<0 \;.
\end{cases}
\end{equation}
Figure \ref{plotfield} illustrates the solution (\ref{ansol}).
More precisely, in the upper plot it is possible to observe how
the magnetic free-energy density is concentrated on the
$\pi$-wall, whereas the lower plot shows that the elastic
free-energy density is mostly localized on the defect. In both
cases, it is possible to notice that the decay pattern from
$\Phi=\pm \frac{\pi}{2}$ on the left $x$-axis to the equilibrium
value $\Phi=0$ becomes constant when we move some magnetic
coherence lengths away from the defect.

We notice that no use of the core region has been made for the
time being. In fact, the differential problem
(\ref{difeq})-(\ref{bcond1}) admits a solution in
$\reals^2\setminus(0,0)$, without any need of excising a finite
region around the defect. We will see below that this will not be
the case when we will have to deal with the derivatives of the
field, and in particular with the dissipation (\ref{dissdef}).

\subsection{Disclination velocity}

We have succeeded in finding a solution of the stationary
Ginzburg-Landau equation for any value of $v$. We will now
complete the eigenvalue analysis, and determine the correct
disclination velocity, by imposing the global dissipation
condition (\ref{dispr}).

We have already noticed that, in the moving reference frame, time
derivatives transform into spatial derivatives. Thus,
\begin{equation}
\mathcal{\dot F}=v\int_\Bs\frac{\partial}{\partial x}
\left(K\big|\nabla \Phi\big|^2+\chi_{\rm
a}H^2\Phi^2\right)dx\,dy=v\left[ \int_{\svreals}\left(K\big|\nabla
\Phi\big|^2+\chi_{\rm
a}H^2\Phi^2\right)dy\right]_{x=-\infty}^{x=+\infty}\;.
\label{fdotcalc}
\end{equation}
When $|x|$ is very large, the principal value of the $q$-integral
in (\ref{ansol}) is dominated by the low $q$ values:
\begin{equation}
\lim_{x\to\pm\infty} \frac{1}{4{\rm i}} \pv\int_{\svreals}
\frac{{\rm e}^{{\rm i}qx-k(q)y}}{q}\,dq = \frac{1}{4{\rm i}}
\lim_{x\to\pm\infty}\pv\int_{\svreals} \frac{{\rm e}^{{\rm
i}qx-y/\xi}}{q}\,dq= \pm\,\frac{\pi}{4} \,{\rm e}^{-y/\xi}\;.
\end{equation}
Thus,
\begin{equation}
\mathcal{\dot F}=-\frac{\pi^2 Kv}{\xi^2}\int_0^{+\infty}{\rm
e}^{-\frac{2y}{\xi}}dy= -\frac{\pi^2 Kv}{2\xi}\;.
\label{fdot}
\end{equation}
The result (\ref{fdot}) admits a simple physical interpretation,
that already exhibits in (\ref{fdotcalc}). The quantity within
square brackets in (\ref{fdotcalc}) is the free-energy contained
in an infinite vertical strip of unit width, centered at $x$. When
$x\to+\infty$, both the elastic and magnetic energy densities
relax to $0$, as Figure \ref{plotfield} shows. As a consequence,
the total energy stored in the right-side strip is negligible. The
picture completely changes in the $x\to-\infty$ limit. A
unit-width strip crossing the $\pi$-wall stores a finite amount of
free energy, measured in (\ref{fdot}). In fact, it is precisely
the difference between the energies stored in those strips that
keeps the defect moving. In the unit time, the defect motion
replaces a left-side strip of width $v$, with free energy given in
(\ref{fdot}), with a right-side strip of equal width, with no free
energy.

While the free energy variation depends only on the asymptotic
structure of the director field, dissipation takes place in the
whole domain. Indeed, it is so strong close to the moving defect,
that we will be forced to exclude the core region in order to
avoid an infinite dissipation. We have:
\begin{align}
\nonumber\mathcal{D}&=\gamma_1 \int_{\svreals^2}
\dot\Phi^2\,dx\,dy=\gamma_1 v^2 \int_{\svreals^2}
\left(\frac{\partial\Phi}{\partial x}\right)^2\,dx\,dy\\
\nonumber&=\frac{\gamma_1 v^2}{16}\int_{\svreals^2}dxdy
\pv\int_{\svreals} dq\pv\int_{\svreals} dq'
{\rm e}^{{\rm i}\big(q+q'\big)x-\big(k(q)+k(q')\big)|y|}\\
\nonumber&=\frac{\gamma_1 v^2}{8}\pv\int_{\svreals}
dq\pv\int_{\svreals} dq'\,
\frac{2\pi\delta\big(q+q'\big)}{k(q)+k(q')}\\
\nonumber&=\frac{\pi\gamma_1 v^2}{2}\int_0^{q_{\rm M}/(2\pi)}
\frac{dq}{\sqrt{q^2 +{\rm i}q\lambda/\xi+1/\xi^2}+\sqrt{q^2
-{\rm i}q\lambda/\xi+1/\xi^2}}\\
&=-\frac{{\rm i}\pi Kv}{2\xi}\int_0^{\xi/\rs} \left(\sqrt{1
+\frac{{\rm i}\lambda}{s}+\frac{1}{s^2}}-\sqrt{1 -\frac{{\rm
i}\lambda}{s}+\frac{1}{s^2}}\right)ds\;,
\label{disscalc}
\end{align}
where $\delta$ denotes the Dirac delta function. The high-$q$
cutoff is needed in order to avoid the logarithmic divergence
which the disclination induces both in the free energy (but not in
its time-derivative) and in the dissipation. This is related to
the inverse of the core radius: $q_{\rm M}=2\pi/\rs$.

When we replace (\ref{fdotcalc}) and (\ref{disscalc}) into the
dissipation principle (\ref{dispr}) we obtain the self-consistency
equation that determines $\lambda$ (\emph{i.e.\/}, $v$), as a
function of the ratio $\xi/\rs$:
\begin{equation}
\int_0^{\xi/\rs} \left(\sqrt{1 +\frac{{\rm
i}\lambda}{s}+\frac{1}{s^2}}-\sqrt{1 -\frac{{\rm
i}\lambda}{s}+\frac{1}{s^2}}\right)ds={\rm i}\pi
\label{lamueq}
\end{equation}
The most interesting region in physical applications is $\xi \gg
\rs $. The integral on the left hand side is dominated by its
logarithmic high-$s$ divergence and we obtain:
\begin{equation}
{\rm i}\lambda\log \frac{\xi}{\rs}={\rm i}\pi
\quad\Longrightarrow\quad v= \frac{2\pi\sqrt{K\chi_{\rm
a}}}{\gamma_1\log \big(\xi /\rs\big)}\;|H| \qquad{\rm when}\quad
\xi \gg \rs\;.
\label{linear}
\end{equation}
However, it is interesting to push the analysis of (\ref{lamueq})
into the opposite regime $\xi\ll \rs $. The low-$s$ terms dominate
the integral on the left hand side and we find:
\begin{equation}
{\rm i}\lambda\,\frac{\xi}{\rs}
 ={\rm i}\pi \quad\Longrightarrow\quad
v  =\frac{2\pi\chi_{\rm a}\rs}{\gamma_1\,}\,H^2 \qquad{\rm
when}\quad \xi  \ll \rs\;.
\label{quadr}
\end{equation}
The quite intriguing quadratic behaviour predicted by
(\ref{quadr}) must be handled carefully. When the external field
becomes so intense that $\xi$ becomes of the order of $\rs$, we
have to question our assumption of $\rs$ being independent from
$\xi$. A more complete theory, that can be derived following the
steps of \cite{03bisl}, would yield an $\rs(\xi)$, and thus a
disclination velocity depending only on the strength of the
applied field.

\begin{figure}[htb]
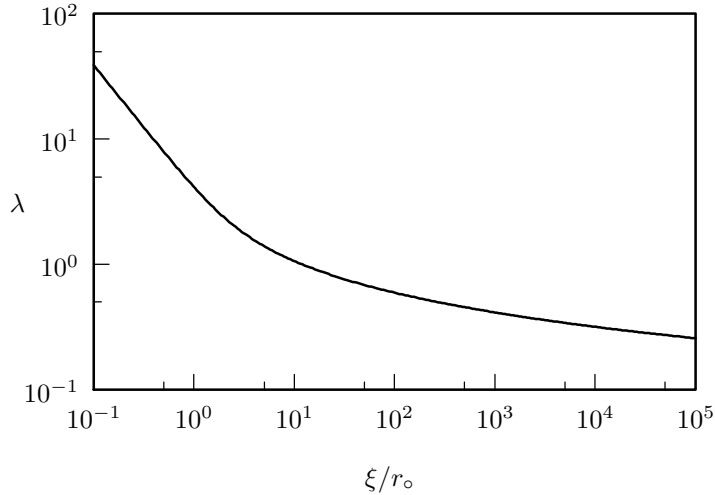

\input fig3.tex
\caption{Disclination velocity as a function of the external
applied field}
\label{lamu}
\end{figure}

Figure \ref{lamu} shows the numerical solution of equation
(\ref{lamueq}). The plot highlights that the transition between
the two asymptotic regimes derived above occurs for $\xi \gtrsim
\rs$. Even though this is the limit up to which we can seriously
trust our analytical result, Figure \ref{lamu} suggests that the
nonlinear effects increase the disclination velocity.

\section{Defect interaction}

Let us now consider two nematic defects of opposite topological
charges $s=\pm \frac{1}{2}$ placed on the $x$-axis at a distance
$\Delta=2d$, but in the absence of any external field. After a
certain period of time these defects amalgamate and thus
annihilate each other. The two cores disappear at once, with a
large and rapid reduction in  free-energy
\cite{91patu,96denn,02svzu}. Imagine, however, that we apply a
magnetic field that favours the director orientation of the
molecules that lie within the defects. The defect speed will be
certainly reduced. In this section we  show that it can be even
reversed. It exists a critical distance such that the defect
interaction is attractive only if their mutual distance is smaller
than the critical one. Otherwise, they repel.

\begin{figure}[htb]
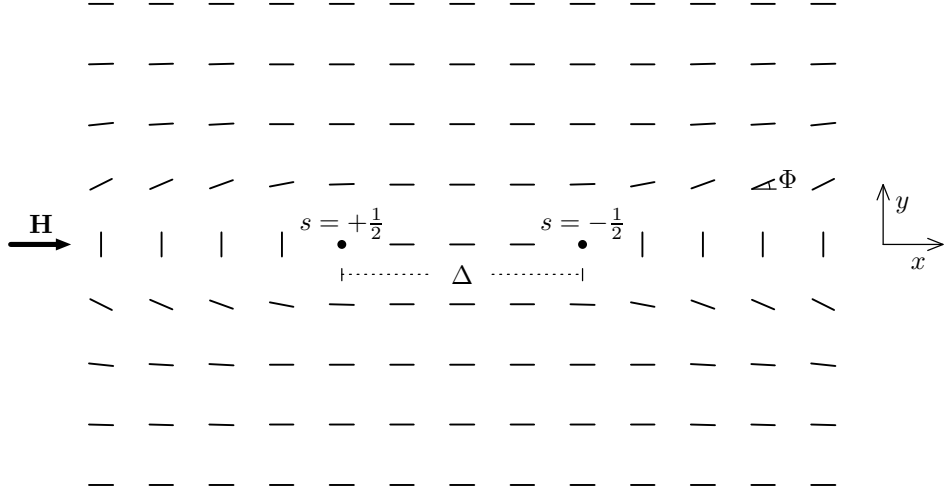

\null\medskip
\input fig4.tex
\caption{Two attracting defects placed at a distance $\Delta$
along the $x$-axis. The external field is parallel both to the
line connecting the defects and to the director orientation
between them.}
\label{2def}
\end{figure}

Throughout this section, we will work out in detail the geometry
illustrated in Figure \ref{2def}. The applied field lies parallel
both to the line connecting the defects, and to the director
orientation between them. In this geometry, the defects may only
approach or separate, so avoiding but more complicated motions
such as mutual rotations.

We compute the speed of two stationarily moving defects. We will
show below that the stationarity assumption holds approximately
even when the defect distance becomes of the order of, or smaller
than, the magnetic coherence length. It must certainly be
abandoned when one wants to describe the complete annihilation
process, and in particular when the defect distance becomes of the
order of the core radius.

In the final subsection we will briefly analyse the case when the
external field determines a generic angle $\alpha$ with respect to
the defect line. In this case the motion may be much more
complicated. However, we are able to estimate how the nature of
the defect interaction depends on $\alpha$. More precisely, we
will determine for which values of $\alpha$ the defect interaction
may become repulsive, and how the critical distance depends on the
external field direction.

The time-dependent Ginzburg-Landau equation (\ref{difeq}) is
linear, due to the parabolic approximation we used for the
magnetic energy. We can thus obtain a solution describing two
stationarily approaching (or separating) defects by simply
superposing two functions of the type (\ref{ansol}). More
precisely, we add a solution describing a defect placed at
$x=+d$, travelling with velocity $-v$, and a solution describing a
defect placed at $x=-d$, travelling with velocity $+v$. The
velocity $v(d)$ will be determined later, again
using the  dissipation principle. The director field is given by:
\begin{align}
\nonumber\Phi(x,y)&=\varepsilon(y)\left[\frac{\pi}{2}\,{\rm
e}^{-\frac{|y|}{\xi}}+\frac{1}{4{\rm i}} \pv\int_{\svreals}
\frac{{\rm e}^{{\rm i}q(x-d)-k(q)|y|}-{\rm e}^{{\rm
i}q(x+d)-k(-q)|y|}}{q}\,dq\right]\\
&=\varepsilon(y)\left[\frac{\pi}{2}\,{\rm
e}^{-\frac{|y|}{\xi}}-\int_0^{+\infty} \cos qx\;\sin\big(qd+\im
k(q)\,|y|\big)\;{\rm e}^{-\re k(q)\,|y|} \;\frac{dq}{q}\right]\;.
\label{sol2def}
\end{align}

\begin{figure}[htb]
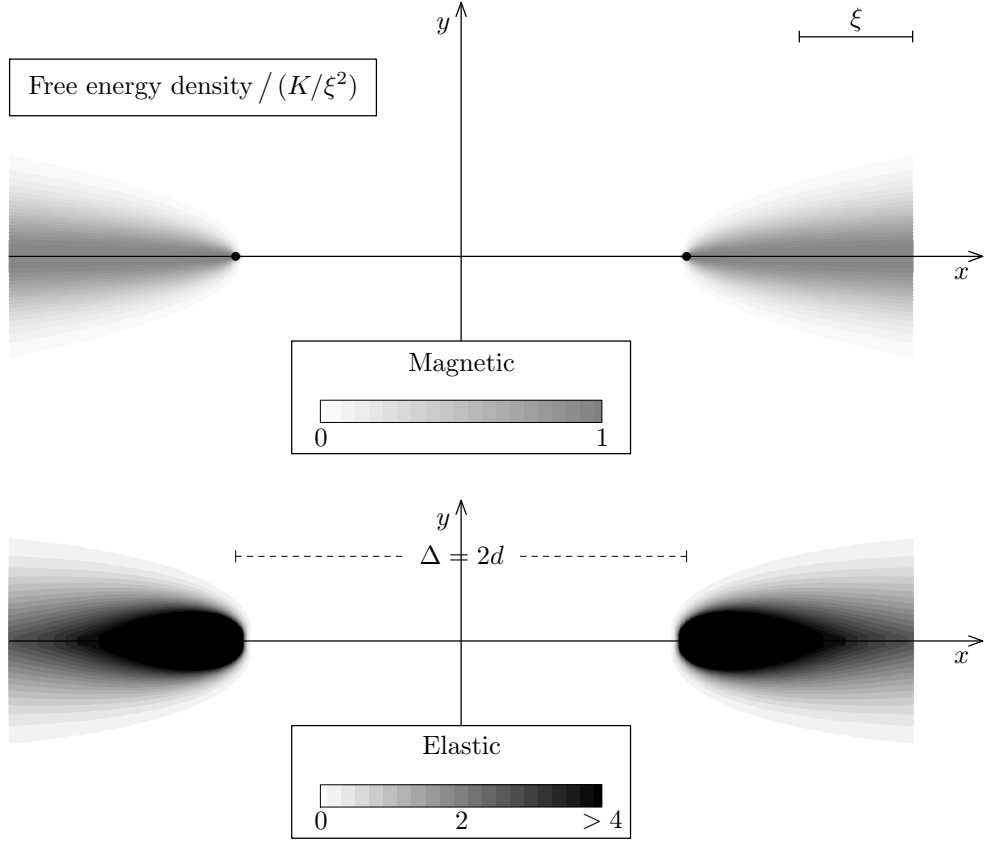

\null\bigskip
\begin{center}
\input fig5a.tex

\medskip

\input fig5b.tex
\end{center}
\caption{Density plot of the magnetic free-energy density (upper
graph) and the elastic free-energy density (lower graph), computed
from the analytical solution (\ref{sol2def}). For these particular
values of the distance and the magnetic coherence length, it will
turn out that the defects are separating (that is, $v<0$). Both
the magnetic and the elastic free-energy densities are mainly
concentrated on the defect walls. Between the defects, the
free-energy density is negligible. }
\label{field2}
\end{figure}

Figure \ref{field2} illustrates the countervailing tendencies of
the elastic and magnetic energies: the elastic contribution aims
at annihilating the defects in order to relax the infinite core
energy. On the contrary, the magnetic field tries to broaden the
intermediate region, where all the molecules are already correctly
aligned.

The free energy associated with the traveling configuration
(\ref{sol2def}) is given by:
\begin{align}
\mathcal{F} (\Delta,\lambda)&=K\pi\left[\frac{\pi L}{\xi}+
\arcsinh\frac{\xi}{\rs}-K_0\left(\frac{\Delta}{\xi}\right)-
\frac{\pi \Delta}{\xi} + 2 \int_0^\infty\frac{\sin^2\dsp\frac
{\Delta s}{2\xi}}{s^2\sqrt{1+s^2}}\;
ds+F_1(\Delta,\lambda)+F_2(\lambda) \right].
\label{fren}
\end{align}
The first term in (\ref{fren}) diverges in the $L\to+\infty$ limit
($L$ is the horizontal scale of the system: $x\in[-L,L]$). This
term corresponds to the energy of the two walls. The second term
is related to the infinite elastic energy stored in the defects
($\rs$ is the core radius as in the preceding section). The next
three terms depend only on the defect distance, but not on their
velocity. In particular, the term containing the modified Bessel
function of the second kind $K_0$ is able to cancel the free
energy divergence when $\Delta\simeq \rs$. Finally, both $F_1$ and
$F_2$ are velocity corrections, which vanish if $\lambda$
(\emph{i.e.\/}, $v$) vanishes:
\begin{align}
\nonumber F_1(\Delta,\lambda)&:=\frac{1}{2}\int_0^\infty
\re\left[\left( \mu(\lambda,s)-\mu(0,s)+\frac{1}{\mu(\lambda,s)}-
\frac{1}{\mu(0,s)}\right)\left(1-{\rm e}^{-{\rm
i}s\Delta/\xi}\right)\right]\frac{ds}{s^2}\\
&-\frac{1}{2}\int_0^\infty \re\left[\left(
\frac{1}{\mu(\lambda,s)}- \frac{1}{\mu(0,s)}\right){\rm e}^{-{\rm
i}s\Delta/\xi}\right]ds \\
\nonumber F_2(\lambda)&:=\int_0^\infty \left(
\frac{1}{\mu(\lambda,s)+\mu(-\lambda,s)}-
\frac{1}{2\mu(0,s)}\right)ds\\
&-\frac{1}{4}\int_0^\infty
\frac{\big(\mu(\lambda,s)-\mu(-\lambda,s)\big)^2}
{\mu(\lambda,s)+\mu(-\lambda,s)}\left(1+\frac{1}{\mu(\lambda,s)
\mu(-\lambda,s)}\right)\frac{ds}{s^2}\;,
\end{align}
where $\mu(s)$ is the positive-real-part square-root of
$\mu^2(s)=1+{\rm i}s\lambda+s^2$, and
$\lambda=\dsp\frac{\gamma_1}{2\sqrt{K\chi_{\rm
a}}}\,\frac{v}{|H|}$ as in the preceding section.

In the absence of backflow, the dissipation stems only from the
director rotation. When computing $\dot\Phi$, we assume that only
the defect position $d$ depends on time. In our stationary
approximation, we thus neglect the time derivative of the velocity
$v$. However, we remark that when we want to determine the
non-stationary effects depending on $\dot v$, we are no longer
allowed to use the solution (\ref{sol2def}). The dissipation
function is given by:
\begin{equation}
\mathcal{D}= \gamma_1 \int_{\svreals^2}\dot\Phi^2\,dx\,dy =
\frac{\pi\dot d^2}{2}\left[
\arcsinh\frac{\xi}{\rs}+K_0\left(\frac{\Delta}{\xi}\right)
+G_1(\Delta, \lambda)+G_2(\lambda)\right]\;,
\label{dissip}
\end{equation}
with
\begin{align}
G_1(\Delta,\lambda)&:=\int_0^\infty \re\left[\left(
\frac{1}{\mu(\lambda,s)}- \frac{1}{\mu(0,s)}\right){\rm e}^{-{\rm
i}s\Delta/\xi}\right]ds \qquad{\rm and}\\
G_2(\lambda)&:=2\int_0^\infty \left(
\frac{1}{\mu(\lambda,s)+\mu(-\lambda,s)}-
\frac{1}{2\mu(0,s)}\right)ds\;.
\end{align}
The functions $G_1$ and $G_2$ vanish in the low-velocity limit
$v\to 0$. The dissipation principle delivers the self-consistency
equation that determines the defect velocity $v=-\dot d\,$:
\begin{align}
&\frac{\partial \mathcal{F}}{\partial d}\;\dot
d+\mathcal{D}=0\qquad\Longleftrightarrow
\label{diss2def}\\
\nonumber &K_1\left(\frac{\Delta}{\xi}\right)+\frac{\Delta }{\xi}
\int_0^\infty\frac{\sin s\ ds}{s\sqrt{\frac{\Delta^2}{\xi^2}+s^2}}
+\xi\,\frac{\partial F_1}{\partial \Delta}-\frac{\lambda}{2}
\left[ \arcsinh\frac{\xi}{\rs}+K_0\left(\frac{\Delta}{\xi}\right)
+G_1(\Delta, \lambda)+G_2(\lambda)\right]=\pi\;.
\end{align}
Before  analyzing  the solutions $\lambda(\Delta)$ of
equation (\ref{diss2def}), we want to stress the importance of the
first two terms in its left-hand side. In fact, if we define
\begin{equation}
f(x):= K_1(x)+ x\int_0^\infty\frac{\sin s}{s\sqrt{x^2+s^2}}\;ds\;,
\label{deffx}
\end{equation}
we have:
\begin{equation}
\left.\frac{\partial
\mathcal{F}}{\partial\Delta}\right|_{v=0}=\frac{K\pi}{\xi}\left[
f\left(\frac{\Delta}{\xi}\right)-\pi\right]\;.\,
\label{defdir}
\end{equation}
Thus, in general, the defects will approach or separate depending
on whether $f(\Delta/\xi)$ exceeds $\pi$ or not. We postpone the
analysis of the properties of $f$ to the next subsection, when we
will generalize equation (\ref{defdir}) to the case of tilted
applied fields. As for now, we only remark that $f(x)=\pi$ when
$x=x_{\rm cr}\doteq 0.377388$. The function $f$ is greater than
$\pi$ (thus inducing defect attraction) when $\Delta<x_{\rm
cr}\,\xi$. Defect repulsion is induced at distances greater than
$x_{\rm cr}\,\xi$.

\begin{figure}[htb]
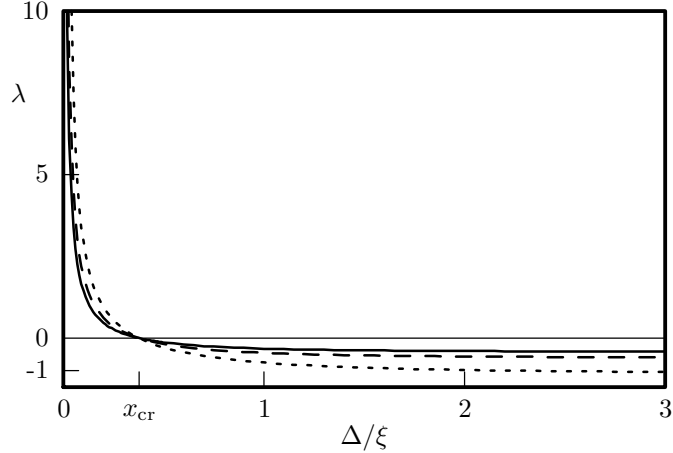

\input fig6.tex
\caption{Defect velocity as a function of the external field for
different values of the core radius $\rs$: $\xi/\rs=10$ (dotted
line), 100 (dashed), 1000 (full). For any value of the core
radius, the defects attract only if their distance is smaller than
$\Delta_{\rm cr}$.}
\label{vel2def}
\end{figure}

Figure \ref{vel2def} illustrates the numerical solutions of
equation (\ref{diss2def}) for three different values of the ratio
between the magnetic coherence length and the core radius. They
exhibit the following properties.
\begin{itemize}
\item In the large distance limit, all Bessel functions decay
exponentially with $\Delta/\xi$. The functions $G_1,G_2$, and the
derivative of $F_1$ vanish too. Furthermore,
\begin{equation}
\lim_{x\to\infty} x\int_0^\infty\frac{\sin
s}{s\sqrt{x^2+s^2}}\;ds=\int_0^\infty\frac{\sin
s}{s}\;ds=\frac{\pi}{2}\;.
\end{equation}
Thus, the large-distance limit of $\lambda$ is given by
\begin{equation}
\lim_{\Delta\to\infty}\lambda (\Delta)=-\frac{\pi}{\arcsinh
(\xi/\rs)}\qquad\Longrightarrow\qquad v\simeq-
\frac{2\pi\sqrt{K\chi_{\rm a}}\;|H|}{\gamma_1\,\arcsinh (\xi/\rs)}
\quad \text{when }\Delta\gg\xi\;.
\label{largedist}
\end{equation}
The defects repel, and move at a constant speed. In fact, if we
compare (\ref{largedist}) with (\ref{linear}), we notice that in
the large-distance limit the defects behave independently, each
moving at the velocity computed in the 1-defect case, since
$\arcsinh (\xi/\rs)\simeq \log (\xi/\rs)$ when $\xi\gg\rs$.

\item The critical distance at which the defect interaction
changes sign does not depend on the core radius $\rs$: all plots
in Figure \ref{vel2def} cross the $x$-axis at $\Delta=x_{\rm
cr}\,\xi$.

\item In the small distance limit, the stationary approximation we
have used is not well justified. In this limit, the velocity
diverges, and all terms depending on $\dot v$ must be taken into
account in the equation of motion. In any case, we remark that the
Bessel function $K_1$ induces a divergence in the derivative of
the free-energy that scales as the inverse of the defect distance.
Figure \ref{vel2def} suggests that the fully nonlinear regime is
limited to distances much smaller than the magnetic coherence
length.
\end{itemize}

\subsection{Tilted external fields}

Let us now imagine that the external field is rotated of an angle
$\alpha$ with respect to the direction of Figure \ref{2def}. In
this case, the nematic director is forced to relax to the external
field direction everywhere but on the topologically irreducible
$\pi$-wall. The motion will be much more complex -- the defects
will rotate, and the $\pi$-wall will not necessarily be straight
at all times. However, it is possible to ascertain whether the
defect interaction will lead to  attraction or  repulsion. To this
aim, we imagine to pin the defects at a distance $\Delta$, and we
look for the stationary director field in the presence of a tilted
external field. Then, we compute the free-energy of the stationary
solution, and we check the sign of its derivative with respect to
the distance. We stress that it is not possible to use this
derivative in a dissipation principle to obtain a defect velocity.
However, its sign will determine whether or not the defects,
whatever their complex motion, will approach each other.

The stationary director field can be simply derived by running
through the above steps again. In the presence of two defects
placed at $x=\pm d$, and an external field tilted of an angle
$\alpha$ with respect to the $x$-axis, the stationary
configuration is given by:
\begin{equation}
\Phi(x,y)= \varepsilon(y)\left[\alpha+
\left(\frac{\pi}{2}-\alpha\right)\,{\rm
e}^{-\frac{|y|}{\xi}}-\int_0^{+\infty} \cos qx\;\sin qd\;{\rm
e}^{-k_0(q)\,|y|} \;\frac{dq}{q}\right]\;,
\label{2deftilt}
\end{equation}
where $k_0(q)$ coincides with the zero-velocity limit of $k(q)$
above: \quad $k_0(q):=\sqrt{q^2+1/\xi^2}\;$.

\begin{figure}[htb]
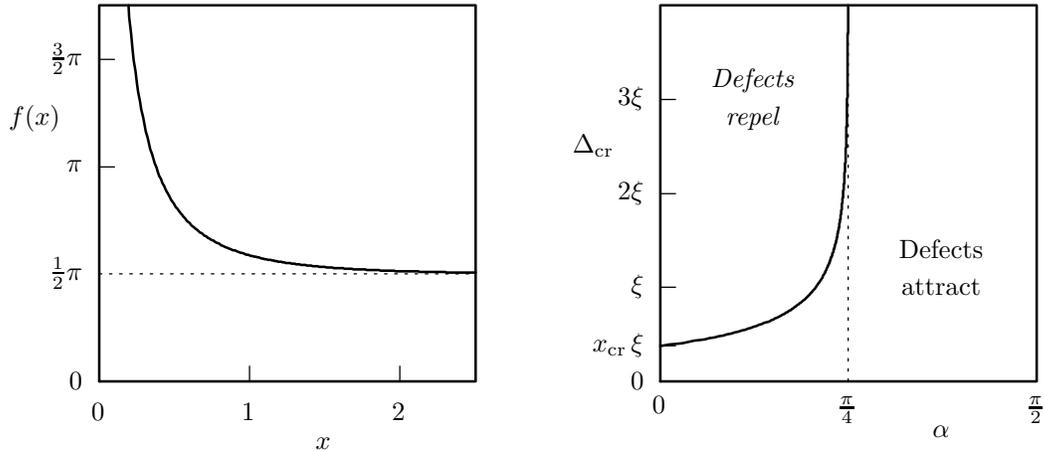

\null\medskip \noindent\null \hfill
\input fig7a.tex
\hfill\hfill
\input fig7b.tex
\hfill
\medskip
\caption{\emph{Left: \/}Plot of the function $f$ defined in the
text. \emph{Right: \/}Critical value of the defect distance as a
function of the tilt angle $\alpha$.}
\label{delcr}
\end{figure}

If we compute the free-energy associated to (\ref{2deftilt}), and
we differentiate it with respect to $\Delta$, we obtain the
generalization of equation (\ref{defdir}):
\begin{equation}
\left.\frac{\partial
\mathcal{F}}{\partial\Delta}\right|_{v=0}=\frac{K\pi}{\xi}\left[
f\left(\frac{\Delta}{\xi}\right)-\big(\pi-2\alpha\big)
\right]\;.\,
\label{defdirtilt}
\end{equation}
with the same $f$ defined in (\ref{deffx}). The left panel of
Figure \ref{delcr} shows the plot of $f$. It enjoys the following
properties:
\begin{itemize}
\item $f'(x)=\frac{1}{2}\,\big(K_0(x)-K_2(x)\big)<0$ for all
$x>0$. Thus, the critical distance at which the defect interaction
becomes repulsive increases when the external field is tilted.
Furthermore, the free energy is a concave function of the distance
between the defects.
\item $\dsp\lim_{x\to 0} f(x)=+\infty$\quad and \quad
$\dsp\lim_{x\to \infty} f(x)=\frac{\pi}{2}\;$. More precisely,
\begin{equation}
f(x)=\frac{1}{x}-\frac{x}{2}\,\log x+O(x)\ {\rm as}\ x\to
0^+\quad{\rm and}\quad f(x)=\frac{\pi}{2}+o\left({\rm
e}^{-x}\right)\ {\rm as}\ x\to +\infty\;.
\end{equation}
Thus, the equation
\begin{equation}
\frac{\partial\mathcal{F}}{\partial\Delta}=0
\quad\Longleftrightarrow\quad f(x) =\pi-2\alpha\;,
\end{equation}
which determines the equilibrium distances, possesses exactly one
solution if and only if $\alpha<\frac{\pi}{4}\,$. This limiting
value for the tilting angle could be easily predicted. In fact, if
the external field determines an angle greater than
$\frac{\pi}{4}$ with the defect line, the director orientation
between the defects costs \emph{more\/} magnetic energy than the
director orientation outside the defects. Thus, in this case, the
external field strengthens the defect attraction at any distance.
\end{itemize}
In summary:
\begin{itemize}
\item When $\alpha=0$, the defects attract
if $\Delta<\Delta_{\rm cr}\big(\frac{\pi}{2}\big)=0.3774\xi$; they
repel when $\Delta>\Delta_{\rm cr}\big(\frac{\pi}{2}\big)$.
\item When $\alpha>0$, the critical distance $\Delta_{\rm
cr}(\beta)$ increases. It diverges when
$\alpha\to\frac{\pi}{4}^-$.
\item When $\alpha\geq\frac{\pi}{4}$ there is no critical
distance: the defects always attract (the external field enhances
the attraction).
\end{itemize}

\section{Discussion}

We have studied the motion of a single disclination, and a
disclination dipole in an external field. Our results show that it
is possible to drive the disclinations by suitably adjusting the
external field direction and strength.

In the case of a single disclination we have shown that the
disclination velocity depends almost linearly on the field strength,
since the coefficient depends on the  logarithm of the
field strength (see equation
(\ref{linear})). Figure \ref{lamu} shows that the linear scaling
is abandoned  when $\xi\simeq \rs$. Equation (\ref{quadr}) shows
that below this regime the disclination velocity is expected to
scale as the square of the applied strength. However, this last
prediction should be tested carefully, since it is surely
influenced by our assumption that the core radius does not depend
on the field strength.

We have also computed the velocity of two opposite disclinations
in the presence of a field that promotes the director orientation
between them. The velocity depends on the defect distance $\Delta$
but, whatever the value of the ratio $\xi/\rs$, they defects
approach if $\Delta$ is smaller that a critical distance
$\Delta_{\rm cr}$; otherwise, they repel. We have finally
generalized our calculations in order to deal with rotated
external fields. Figure \ref{delcr} (right) shows how the
$\Delta_{\rm cr}$ depends on the angle $\alpha$ that the external
field determines with the lie which connects the defects.

The introduction of backflow may change the picture we have
developed, both in the one and two-disclination cases. In the one
disclination case the dissipation may sometimes be significantly
reduced. The system can match the disclination motion with a flow
that almost cancels the dissipation in the crucial core region.
Whether this occurs seems likely to depend on the charge of the
central disclination. We note that central physical and
mathematical issues associated with backflow in nematic liquid
crystal problems still remain open even over a quarter of a
century after its essential role was first realized \cite{78clle}.

Likewise in the two-disclination problem, our computed defect
velocity is symmetrical in both
disclinations. This is also an effect of the no-backflow
simplification. A generalization of the present study should allow
to compute the different approaching or separating velocities. It
could even happen that one  defect moves towards the other, but that
the other retreats faster still, allowing the defects eventually
to separate. However, the existence of a critical
distance that reverses the defect interaction can not be erased by
backflow effects. The critical distance stems from the balance
between the elastic and magnetic free-energy gains. The magnetic
gain does not depend on the defect distance, while the elastic
gain vanishes when the defects move apart. Thus, at some
intermediate distance one overwhelms the other.

\section*{Acknowledgements}

This work has been supported by a British Council Anglo-Italian
collaborative grant. P.B.\ acknowledges the hospitality of
the University of Southampton, where this work
was carried out.

\end{document}